\documentclass[
prl,
twocolumn,
aps,
]{revtex4}
\usepackage{amsmath,amssymb,amsfonts}
\usepackage{graphicx}

\begin{document}
\title{Anomalous Diffusion}
\author{Mendeli H. Vainstein}
\email[E-mail: ]{mendeli@fis.unb.br}
\affiliation{Institute of Physics and International Center of Condensed Matter Physics, University of Bras\'{\i}lia, P.O. Box: 04513, 70919-970 Bras\'{\i}lia-DF, Brazil}
\author{Luciano C. Lapas}
\email[E-mail: ]{luciano@fis.unb.br}
\affiliation{Institute of Physics and International Center of Condensed Matter Physics, University of Bras\'{\i}lia, P.O. Box: 04513, 70919-970 Bras\'{\i}lia-DF, Brazil}
\author{Fernando A. Oliveira}
\email[E-mail: ]{fao@fis.unb.br}
\affiliation{Institute of Physics and International Center of Condensed Matter Physics, University of Bras\'{\i}lia, P.O. Box: 04513, 70919-970 Bras\'{\i}lia-DF, Brazil}
\begin{abstract}
Recent investigations  call attention to the dynamics of anomalous diffusion and its connection with basic principles of statistical mechanics. We present here a short review of those ideas and their implications. 
\end{abstract}

\maketitle

1. Introduction. - Diffusion is a fundamental problem in statistical physics~\cite{Brown28,Einstein05,Smoluchowski06,Langevin08} and much more simple to describe than reaction rates~\cite{Kramers40,Hanggi90,Oliveira98},  for example. From the studies of diffusion, it is possible to obtain a direct description of important concepts in physics, such as the ergodic hypothesis (EH)~\cite{Lee07a,Khinchin49,Boltzmann74,Lee01} and the fluctuation-dissipation theorem (FDT)~\cite{Langevin08,Kubo66,Costa03}. Some of the first concepts of statistical mechanics, for instance the Gaussian distribution of particles, were obtained in the study of diffusion, which also led promptly to other concepts as the Boltzmann equilibrium distribution.  In our study of diffusion, we use the Mori formalism~\cite{Mori65} which has a well defined memory function for non-Markovian systems.  In this context, the Kubo response function~\cite{Zwanzgi01}  is of special importance, since it was formulated in terms of correlation functions which can be obtained by the use of linear response theory and  are directly connected with experiments, such as light or neutron scattering~\cite{Oliveira81}.
In this article we display some recent results in this area, and we point the reader to selected literature directly related to this topic.	

2. Anomalous Diffusion. -  Anomalies in the traditional Einstein diffusion have been the focus of extensive research in many disciplines~\cite{Astumian02,Morgado02,Bulashenko02,Costa03,Bao03,Budini04,Dorea06,Bao06}. In order to address this problem let us consider the generalized Langevin equation (GLE) in the form~\cite{Mori65,Kubo66,Zwanzgi01}
 \begin{equation}
\frac{dP (t) }{dt}=-\int_{0}^{t}\Pi ( t-t^{\prime
}) P( t^{\prime }) dt^{\prime }+F(t) ,
\label{GLE}
\end{equation}
 where $P$ is the particle momentum, $\Pi (t) $ is a kernel, or memory function, and $F(t) $ is a random force, which fulfills $\langle F(t)\rangle =0$ and the fluctuation-dissipation theorem (FDT):
\begin{equation}
\langle F(t) F( t^{\prime }) \rangle
=\langle P^{2}\rangle _{eq}\Pi ( t-t^{\prime }).
\label{FDT}
\end{equation}
The usual manner to study the diffusive dynamics is to investigate the mean square displacement of the particles, given by
\begin{equation}
\lim_{t \rightarrow \infty} \langle x^{2}(t)\rangle \propto t^{\alpha }.
\label{x2}
\end{equation}
The exponent $\alpha $ classifies the type of diffusion: for $\alpha =1$, we have normal diffusion; for $0<\alpha <1$, subdiffusion; and $\alpha >1$, superdiffusion. To obtain average values, it is important to define the correlation function  for the dynamical operator $P$ as
\begin{equation}
C_P(t) = \langle P(t)  P(0) \rangle,
  \label{CP}
\end{equation}
where the brackets $\left\langle \right\rangle$ denote an ensemble average. One can also define the normalized correlation function
\begin{equation}
R(t)= C_P(t)/C_P(0),
\end{equation}
which obeys the equation
\begin{equation}
\frac{dR(t)}{dt}=-\int_{0}^{t}\Pi (t-t^{\prime}) R( t^{\prime }) dt^{\prime }.
\label{R}
\end{equation}
Its  Laplace transform yields
 \begin{equation}
 \tilde{R}(z) =\frac{1}{ z+\tilde{\Pi}(z) }.\label{Rz}
\end{equation}
Recently, an important achievement  in the dynamics of diffusion  has been obtained by Morgado \emph{et al.}~\cite{Morgado02}. They have shown that if
 \begin{equation}
 \lim_{z\rightarrow 0} \tilde{\Pi}(z) \approx cz^{\nu }, \label{Piz}
  \end{equation}
 where $c$ is a positive nondimensional constant, then
 \begin{equation}
 \alpha =1+\nu. \label{alpha}
 \end{equation}
 Consequently, the  behavior of the memory $\tilde{\Pi}(z)$ for small $z$ determines the long range behavior of the diffusion, Eq.~(\ref{x2}). This result has found many applications: the study of molecular motors~\cite{Bao03,{Bao06}}, anomalous diffusion~\cite{Costa03,Dorea06,Vainstein06,Lapas07}, and the dynamics of dipolar chains~\cite{Toussaint04} are a few examples.

In a subsequent work, Costa \emph{et al.}~\cite{Costa03} have shown that the average value of the momentum is
\begin{equation}
\langle P^{2}(t)\rangle =\langle P^{2}\rangle _{eq}+R^{2}(t) [ \langle
P^{2}(0)\rangle - \langle P^{2}\rangle _{eq} ].
\label{P2}
\end{equation}
 In this equation the average value will be the equilibrium value if
\begin{equation}
\lim_{t\rightarrow \infty }R(t)=0,
 \label{Ir}
\end{equation}
  i. e.,  irreversibility is a necessary and sufficient condition for ergodicity to hold in diffusion. For systems which   violate the condition  given in Eq.~(\ref{Ir}), there will be no ergodicity. Using the final value theorem and Eq.~(\ref{Rz}) we obtain
\begin{equation}
\lim_{t\rightarrow \infty }R(t) =\lim_{z\rightarrow 0}z\tilde{R}(z)= 
\lim_{z\rightarrow 0}\frac{1}{1+cz^{\alpha-2}}.  
\label{final}
\end{equation}
This relation shows that for most of the diffusive regimes, $0 < \alpha < 2$, the equilibrium condition, Eq.~(\ref{Ir}), holds. However, for  ballistic diffusion, $\alpha=2$, it is not fulfilled
\begin{equation}
\lim_{t\rightarrow \infty }R(t) =\lim_{z\rightarrow 0}\frac{1}{1+c} \neq 0\text{.}
  \label{bal}
\end{equation}
This means that ballistic diffusion violates ergodicity and the fluctuation dissipation theorem~\cite{Costa03}.  In other words, if the ballistic system is not initially equilibrated, then it will never reach equilibrium and the final result of any measurement will depend on the initial conditions. In this situation the EH will not be valid. For $\alpha >2$,  no memory of the initial condition is lost ($R( t \rightarrow \infty )=1$) and the process is not diffusive, being an activated process for which the GLE  does not work~\cite{Costa03}.
The  results of this letter apply to all kinds of diffusion, $0 < \alpha < 2$, described by a GLE independent of the memory range. This gives origin to new studies in ballistic diffusion~\cite{Bao03,Bao06,Lapas07}. 

For any initial distribution of values $P(0)$, with $\langle F(t) P(0)\rangle =0$, it is possible to obtain the temporal evolution of the moments of $P$,
\begin{equation}
\langle P(t) \rangle =\langle P(0) \rangle R(t),
\label{P}
\end{equation}
Again, after an infinite time, we expect that $ \langle P(t \rightarrow \infty ) \rangle =0$. However, for ballistic diffusion, this is not the case,  as can be seen from Eq.~(\ref{bal}). The astonishing result here is the existence of a residual current such as in superfluids.

3. Nonexponential behavior. -  From the above results, it is quite clear that the correlation for ballistic diffusion will not decay exponentially to equilibrium. Besides the ballistic case, any anomalous regime will present nonexponential decay. Even for normal diffusion, a large number of relaxations may be nonexponential~\cite{Morgado02}.
There are a large number of phenomena where the systems do not relax immediately to equilibrium. Those phenomena, usually  associated with non-aging, have nonexponential relaxation and are most commonly described by power  laws or stretched exponentials.  The study of anomalous relaxation has produced  quite interesting results~\cite{Kauzmann48,Parisi97,Ricci00,Metzler00,Metzler04,Holek01,Rubi03,Hentschel07}.

For a system described by a GLE of the form Eq.~(\ref{GLE}), the evolution relies on the noise that drives the particles. For a harmonic noise~\cite{Morgado02}
\begin{equation}
F(t)=\frac{1}{\sqrt{2k_BT}}\int  \sqrt{\rho(\omega)}\cos(\omega t+\phi(\omega))d\omega,
\label{noise}
\end{equation}
where $\rho$ is the noise spectral density, $k_B$ is the Boltzmann constant, and $\phi$ is a set of random phases in the range  $0\leq \phi \leq 2\pi$.
A systematic study carried on by Vainstein \emph{et al.}~\cite{Vainstein06} has shown that the spectral density plays a fundamental role in the description of stochastic processes as we shall see.
First, the memory function $\Pi(t)$ can be easily obtained by the use of the FDT, Eq.~(\ref{FDT}), as
\begin{equation}
\Pi(t) = \int  \rho(\omega)\cos(\omega t)d\omega,
\end{equation}
 in such a way that the average cancels the random terms and obviously the memory is a deterministic even function. The Laplace transform  of the memory $\tilde{\Pi}(z)$ is an odd function in $z$; therefore,  $\tilde{R}(z)$ given in  Eq.~(\ref{Rz}) is also an odd function in $z$. This implies that by inverting the Laplace transform, $R(t)$ is an even function of $t$
\begin{equation}
R(-t) = R(t) \label{even}.
\end{equation}

 Second, the condition given by Eq.~(\ref{Piz}) to determine the exponent $\alpha$, Eq~(\ref{alpha}), for a spectral density of the form
 \begin{equation}
  \lim_{\omega \rightarrow 0} {\rho}(\omega ) \approx \omega^{\beta}, \label{ro}
  \end{equation}
becomes
\begin{equation}
\nu =
\left \{
  \begin{array}{ll}
   \beta, & \mbox{  $\beta < 1$};\\
   \\
    1, & \mbox{  $\beta \geq 1$}. \label{beta}
  \end{array}
 \right.
 \end{equation}
This shows that a noise spectral density in the form of a power law can produce only diffusive motion in the region $0\leq \alpha \leq 2$. Diffusive motion beyond ballistic is not allowed, what can be observed by using the Laplace transform.

 To study relaxation we need to know $R(t)$, which can be calculated analytically in restricted cases, being obtained  numerically most of the times.  A recent study shows that exponential decay, power laws or even Mittag-Leffler functions are particular cases of  a more general function~\cite{Vainstein06} which approximates the decay. Indeed, considering time reversal symmetry~\cite{Oliveira81} and the discussion above (see Eq.~(\ref{even})), the correlation function must be even and cannot be any of those forms. We shall expose here the conditions under which it is possible to obtain an approximately exponential decay, what happens in certain circumstances for normal diffusion. In this case, 
\begin{equation}
 \gamma =  \lim_{z \rightarrow 0}\tilde{\Pi}(z) = \frac{\pi}{2}\rho(0).
\end{equation}
Consequently, the friction in the usual Langevin equation is nothing more than the noise spectral density for the lower modes. For an arbitrary memory, the system has a rich behavior; even for normal diffusion it is possible to show the existence of at least three time ranges~\cite{Vainstein06}.  A normal diffusion can be obtained using a spectral density of the form~\cite{Morgado02}
\begin{equation}
\rho(\omega) =
\left \{
  \begin{array}{ll}
   \frac{2\gamma}{\pi}, & \mbox{  $\omega < \omega_{s}$};\\
   \\
    0, & \mbox{ $\omega > \omega_{s}$} . \label{beta1}
  \end{array}
 \right.
 \end{equation}
For a broad band noise spectral density, $\gamma / \omega_{s} < 1$, and long times $t > \gamma^{-1}$ it is possible to decouple Eq.~(\ref{R}) to obtain an exponential decay of the form  $R(t)=\exp{(-\gamma t)}$, which will bring the system to equilibrium, that is,   $R(t\rightarrow\infty ) \rightarrow 0$.

Most of the experimental situations where anomalous relaxation is present arise in complex, nonlinear or far from equilibrium structures in which detailed balance does not hold. Good examples can be found in supercooled liquids~\cite{Kauzmann48} and  in glasses~\cite{Parisi97,Ricci00}. Those systems, however, apparently do not have any easy analytical solution.  On the another hand, diffusion can present closed solution for the main expectation values, and arises as a simple laboratory for the discussion of  those properties.

4.  Nonlinear dynamics I: Chain Dynamics - As mentioned before~\cite{Toussaint04}, chain dynamics is a quite interesting subject where the application of those ideas may lead to important results. There is a particular problem that has attracted our attention: the breaking process of a chain of $N$ monomers of mass $M$ subject to a strain $S$.
The equation of motion for monomer $j$, $j=1,2,..N$, is~\cite{Oliveira94,Oliveira98b}
\begin{equation}
M\frac{d^2X_j }{dt^2}= G_j - G_{j+1} -M \gamma \frac{dX_j}{dt}+F(t),
\label{LE}
\end{equation}
where the forces 
\begin{equation}
G_j=-\frac{\partial U(a+S+y)}{\partial y}|_{X_j-X_{j-1}}
\label{FLJ}
\end{equation}
 are derived from the interparticle potential $U$. 
In this situation, it is simpler to treat the random force $F(t)$ as delta correlated, i.e.,  $\Pi(t)=2\gamma\delta(t)$. We then define an effective potential and consider the system as a one body Kramers problem. However, simulations show that the breaking rate is around a hundred times smaller than the usual Kramers rate~\cite{Kramers40}. Several approaches were used in order to overcome the problem; one of them was to consider that the collective motion of the chain generates a harmonic noise, which is correlated and has an associated memory. Those were to be added to Eq.~(\ref{LE}) in order to consider a non-Markovian analysis~\cite{Oliveira95}. This improves the results,  but does not solve the problem. This is an important issue, because the simulation really describes a very important experiment. Polymers used as an additive in a turbulent flow have breaking rates which are up to $10^{-6}$ smaller than those computed using the simple Kramers theory; the simulation~\cite{Maroja01}, on the other hand, is in perfect agreement with the experiments.

5.  Nonlinear dynamics II:  Synchronization -  Nonlinear dynamics plus noise is an explosive mixture with a large number of unexpected results. The study of the evolution of maps subject to a common noise shows different examples of  synchronization~\cite{Fahy92,Longa96}. Besides that, the study of Langevin trajectories shows very nice patterns both for systems  without memory~\cite{Ciesla01} and with memory~\cite{Morgado07}. Synchronization  of many different phenomena arises continuously in the literature~\cite{Acebron05}.

6. Final remarks. -  Diffusion, one of the  simplest phenomena in physics, is a starting point for the study of simple and complex fluids.  Many theorems in statistical physics and even  proper applications  of its formalism, such as a ``simple linear response theory", rely on the  ergodic hypothesis  and its validation. Again, diffusion shows up as a simple way to address the problem.  In this work we have discussed various regimes of  anomalous diffusion, which are ergodic in the range of exponents $0 < \alpha < 2$, where $\alpha$ defines  the asymptotic behavior of the diffusion, Eq.~(\ref{x2}). For $\alpha=2$, we have the special ballistic case, for which ergodicity is not valid, as we have seen. In recent years, molecular motors have been receiving  a lot of attention~\cite{Bao03,Bao06} because of their large potential for pure and applied science. In this subject, a discussion such as the one presented in this paper may be very useful. Moreover, there are many situations  that present violation of the EH, particularly in glassy systems~\cite{Parisi97,Ricci00}, and others were the EH holds. For example, dynamical simulations and equilibrium statistical mechanics were recently used to treat glass transition calculations~\cite{Hentschel07}. The agreement found between the two approaches is a strong indication of the validity of the EH. This is a quite surprising result for such a complex system. Disordered systems are still a large universe in which to explore the basic assumptions of statistical mechanics.

\acknowledgments
 This work was supported by Coordena\c{c}\~ao de Pessoal de N\'{\i}vel Superior (CAPES), Conselho Nacional de Desenvolmento Cient\'{\i}fico (CNPq), Financiadora de Estudos e Projetos (FINEP) and Funda\c{c}\~ao de Amparo \`a Pesquisa do Distrito Federal (FAPDF).

\bibliographystyle{unsrt}

\end{document}